\documentclass[twocolumn,12pt]{article}
\usepackage[body={6.5in,9in}]{geometry}

\title{Mathematics, Biology, and Physics: Interactions and Interdependence}
\author{Michael C. Mackey\thanks{e-mail: mackey@cnd.mcgill.ca,
Departments of Physiology, Physics \& Mathematics and Centre for
Nonlinear Dynamics, McGill University, 3655 Promenade Sir William
Osler, H3G 1Y6 Montreal, QC, CANADA} \and Mois\'es
Santill\'{a}n\thanks{e-mail: moyo@esfm.ipn.mx, Permanent address:
Depto. de F\'\i sica, Esc. Sup. de F\'\i sica y Matem\'aticas,
Inst. Polit\'ecnico Nal. 07738 M\'exico D. F., M\'EXICO}}
\date{\today}

\begin{document}
\maketitle

\section*{Introduction}

Modern science  means that which has a solid conceptual framework,
and which considers experimental results as the ultimate litmus
test against which to validate any theoretical construct. Its
birth can be traced back to the 16th and 17th Centuries. The work
of people like Nicholaus Copernicus, Galileo Galilei, Johannes
Kepler, William Harvey, Vesalius, and others was seminal to this
development. Before the so-called Scientific Revolution, natural
philosophers (the forefathers of scientists as we know them today)
did not perform experiments as manual labor was considered a lower
class activity. This attitude, inherited from the Greeks, changed
between the 16th and the 18th Centuries, as merchants and
craftsmen gained economic and political power. As a result
economics, politics, and science went through significant changes.
In that period democracy, capitalism, and modern science were
founded and emerged as the cornerstones of a new era.

During the Enlightenment,  in the latter part of the 18th and
early part of the 19th Centuries, scientific disciplines started
to be hierarchically classified. This classification works well in
some instances, and without it dealing with the rapidly growing
body of knowledge of the past 200 years would have been difficult.
However, it fails to fairly represent the interdisciplinary work
which has been, and continues to be, highly important. In this
paper we give a taste of the rich historical relation between
physics, mathematics, and the biological sciences. We argue that
this will continue to play a very important role in the future,
based on historical examples and on a brief review of the current
situation.

\section*{The 18th and 19th Centuries}

Electrophysiology is the science that studies the interaction
between electromagnetic fields and biological tissues. This
includes the generation of electric or magnetic fields and
electric currents in some specialized organs, the intrinsic
electric and magnetic properties of tissue, the response of
specialized cells (like neurons and muscle cells) to stimulation,
etc. Up to the middle of the 19th Century, the historical
development of electrophysiology paralleled that of
electromagnetism. The first electric generating machines and the
Leyden jar were constructed to produce static electricity for a
specific purpose: to ``electrify" and to stimulate humans. The
Voltaic pile was developed with the idea of galvanic (i.e. direct
current, as opposed to faradic or alternating current)
stimulation. Bioelectric and biomagnetic measurements were the
incentive for the development of sensitive measurement
instruments, like the galvanometer and the capillary electrometer.
Thus, it is no surprise that some scientists of the time made
important contributions to the development of both the biological
and the physical sciences. In the following paragraphs we present
a brief review of the work of some of these interdisciplinary
workers. We do not attempt to present a detailed review of the
history of electrophysiology, as our purpose is only to exemplify
the rich interdisciplinary interactions of the 18th and 19th
Centuries.

The essential invention necessary for the application of a
stimulating electric current was the Leyden jar (a capacitor
formed by a glass bottle covered with metal foil on the inner and
outer surfaces), independently invented in Germany (1745) and The
Netherlands (1746). With it, Benjamin Franklin's experiments
allowed him to deduce the concept of positive and negative
electricity in 1747. Franklin also studied atmospheric electricity
with his famous kite experiment in 1752 (many American school
children have heard the apocryphal stories of Franklin flying
kites during thunderstorms strings soaked in salt water).

The most famous experiments in neuromuscular stimulation of the
time were performed by Luigi Galvani, professor of anatomy at the
University of Bologna. His first important finding is dated
January 26, 1781. A dissected and prepared frog was lying on the
same table as an electric machine. When his assistant touched the
femoral nerve of the frog with a scalpel, sparks were
simultaneously discharged in the nearby electric machine, and
violent muscular contractions occurred. (It has been suggested
that the assistant was Galvani's wife Lucia, who is known to have
helped him with his experiments). This is cited as the first
documented experiment in neuromuscular electric stimulation.

Galvani continued the stimulation studies with atmospheric
electricity on a prepared frog leg. He connected an electric
conductor between the side of the house and the nerve innervating
the frog leg. Then he grounded the muscle with another conductor
in an adjacent well. Contractions were obtained simultaneous with
the occurrence of lightning flashes. In September 1786, Galvani
was trying to obtain contractions from atmospheric electricity
during calm weather. He suspended frog preparations from an iron
railing in his garden by brass hooks inserted through the spinal
cord. Galvani happened to press the hook against the railing when
the leg was also in contact with it. Observing frequent
contractions, he repeated the experiment in a closed room. He
placed the frog leg on an iron plate and pressed the brass hook
against the plate, and muscular contractions occurred.
Systematically continuing these experiments, Galvani found that
when the nerve and the muscle of a frog were simultaneously
touched with a bimetallic strip of copper and zinc, a contraction
of the muscle was produced. This experiment is often cited as the
classic study to demonstrate the existence of Òanimal electricity.
Galvani did not understand the mechanism of the stimulation with
the bimetallic strip. His explanation for this phenomenon was that
the bimetallic strip was discharging the animal electricity
existing in the body.

Galvani's investigations intrigued his friend and colleague
Alessandro  Volta (professor of physics in Pavia), who eventually
came up with a totally different (and correct) explanation for the
phenomena that Galvani was trying to explain. In the process,
Galvani and Volta maintained their friendship (in spite of their
differences of scientific opinion), and Volta developed the ideas
that eventually led to the invention of the Voltaic pile in 1800
(forerunner of the modern battery), a battery that could produce
continuous electric current. Incidentally, Volta only completed
the equivalent of his doctoral dissertation when he was 50 years
old!

All of these  contributions to electrophysiology were
experimental. The first significant theoretical contributions were
made by the German scientist and philosopher Hermann Ludwig
Ferdinand von Helmholtz. A physician by education and, in 1849,
appointed professor of physiology at K\"onigsberg, he moved to the
chair of physiology at Bonn in 1855 and, in 1871, was awarded the
chair of physics at the University of Berlin. Helmholtz's
fundamental experimental and theoretical scientific contributions
in the field of electrophysiology included the demonstration that
axons are extensions of the nerve cell body, the establishment of
the law of conservation of energy (the First Law of
Thermodynamics), the invention of the myograph, and the first
measurement of the action potential conduction velocity in a motor
nerve axon. Besides these, the contributions of Helmholtz to other
fields of science include fundamental work in physiology,
acoustics, optics, electrodynamics, thermodynamics, and
meteorology. He invented the ophthalmoscope and was the author of
the theory of hearing from which all modern theories of resonance
are derived. Another important contribution to the development of
biophysics was Helmholtz's philosophical position in favor of
founding physiology completely on the principles of physics and
chemistry at a time when physiological explanations were based on
vital forces which were not physical in nature.

\section*{The 20th Century}

In the 18th and 19th Centuries  interdisciplinary research
bridging physics, mathematics and biology was carried out by
scientists educated as physicians. The 20th Century witnessed a
reversal of this trend with major contributions to biology from
people with solid backgrounds in physics and mathematics. There
are two of these disciplines in which the contributions by
physicists and mathematicians were particularly important:
electrophysiology (following the tradition of Galvani, Volta,
Helmholtz, etc.) and molecular biology.

\subsection*{Electrophysiology}

The growth of biophysics owes  much to A. V. Hill, whose work on
muscle calorimetry was essential to our understanding of the
physiology of muscle contraction. Hill received an undergraduate
degree in physics and mathematics, and a doctorate in physiology,
all from Cambridge. Besides his work on muscle contraction, Hill
also addressed problems related to the propagation of the nervous
impulse, the binding of oxygen by hemoglobin, and on calorimetry
of animals. He discovered that heat is produced during the nerve
impulse. Hill's original papers reveal an elegant mixture of
biological concepts and experiments together with physical and
mathematical theory and insight. His discoveries concerning the
production of heat in muscle earned him the Nobel Prize in 1922,
and his research gave rise to an enthusiastic following in the
field of biophysics. He was instrumental in establishing an
extremely successful interdisciplinary school in Cambridge, whose
investigators received a number of Nobel prizes.

A few years later Bernard Katz,  working at University College
London with his student Paul Fatt, made a major advance in our
understanding of the chemical and quantal nature of synaptic
transmission in the papers ``An analysis of the end-plate
potential recorded with an intra-cellular electrode" and
``Spontaneous subthreshold activity at motor nerve endings", which
were marvels of experimental investigation combined with
mathematical modelling of stochastic processes. Katz was one of
the recipients of the 1970 Nobel Prize for ``discoveries
concerning the humoral transmittors in the nerve terminals and the
mechanism for their storage, release and inactivation."

Jumping back a few decades, the German physical chemist  Walter
Nernst was interested in the transport of electrical charge in
electrolyte solutions. His work intrigued another physicist, Max
Planck, one of the fathers of modern quantum theory, who extended
Nernst's experimental and theoretical work, eventually writing
down a transport equation describing the current flow in an
electrolyte under the combined action of an electric field and a
concentration gradient. This work lay largely forgotten until the
1930s, when it was picked up by the physicist Kenneth S. Cole at
Columbia University and his graduate student David Goldman
(originally trained in physics). They realized that the work of
Nernst and Planck (in the form of the Nernst-Planck equation)
could be used to describe ion transport through biological
membranes and did so with great effect. Their work resulted in the
development of the Goldman equation, which describes the membrane
equilibrium potential in terms of intra- and extracellular ionic
concentrations and ionic permeabilities. This background
theoretical work of Nernst and Planck was also instrumental in
helping Cole to experimentally demonstrate that there was a
massive increase in membrane conductance during an action
potential.

Two of the most distinguished alumni  of Hill's Cambridge
interdisciplinary school were A. L. Hodgkin and A. F. Huxley. Both
studied physics, mathematics, and physiology at Trinity College,
Cambridge, whose high table included, at that time, an astonishing
array of scientific talent with people like J. J. Thomson, Lord
Rutherford, F.W. Aston, A.S. Eddington, F.G. Hopkins, G. H. Hardy,
F.J.W. Roughton, W.A.H. Rushton, A.V. Hill, and E.D. Adrian.
Hodgkin and Huxley developed a long-lasting collaboration,
interrupted only by the outbreak of World War II.

In 1938 Hodgkin spent the summer with  Cole at Woods Hole, and
they demonstrated the overshoot of the action potential which had
significant implications in terms of potential ionic mechanisms.
It seems reasonable to suppose that Cole and Hodgkin discussed the
possible meanings of these discoveries and what types of
experiments were needed to determine exactly what was going on.
Because of their training they would have seen that some means
must be found to bring under experimental control the variable
(either membrane current or membrane voltage) that is responsible
for the all-or-nothing behavior of the action potential. Hence
taming the action potential required either controlling the
current or the voltage. They undoubtedly realized that space
clamping was necessary for both current and voltage clamping and
since both knew cable theory, they knew that space clamping was
best done by drastically reducing internal resistance (space
clamping) so the space constant was much longer than the length of
axon under study.

The Second World War interrupted these  investigations and Cole,
like hundreds of other scientists, was caught up in the war
effort. Cole moved from Columbia to the Manhattan Project in
Chicago and worked on radiation dosimetry and radiation damage in
tissues during the war. After the war he was at the University of
Chicago for a few years. When the war was over, one of the
positive outcomes was the existence of high-input impedance vacuum
tubes that had been developed for the amplifiers in radar
receivers. Cole, working with Marmont in Chicago, used these new
electronic advances to build a feedback circuit that allowed them
to space clamp axons. These axons developed an all-or-none action
potential when sufficiently depolarized, and the implication was
that voltage clamping was necessary to tame the axon to measure
the dependence of membrane current on membrane voltage.

Shortly after the war (1948), Hodgkin  visited the United States
and Cole's laboratory in Chicago, and realized that the results of
the space clamp experiments meant that voltage clamping was the
way to go. On his return to England he teamed up with Huxley to
really measure what was going on during the generation of an
action potential in the squid giant axon. This work was published
in a brilliant series of five papers in the Journal of Physiology
in 1952. The final one is an intellectual tour de force combining
both experimental data analysis and mathematical modelling (the
Hodgkin-Huxley equations) that eventually won Hodgkin and Huxley
the Nobel Prize in 1963, along with J.C. Eccles, ``for their
discoveries concerning the ionic mechanisms involved in excitation
and inhibition in the peripheral and central portions of the nerve
cell membrane."  Huxley the mathematician/physiologist was not
content to stop there, however, and went on to publish his
celebrated review of muscle contraction data and its synthesis
into the mathematically formulated cross bridge theory in 1957, a
theory that still stands in its essential ingredients today.

The Hodgkin-Huxley  model for excitability in the membrane of the
squid giant axon is complicated and consists of one nonlinear
partial differential equation coupled to three ordinary
differential equations. In the early 1960s Richard FitzHugh
applied some of the techniques that he had learned from the
Russian applied mathematics literature to an analysis of the
Hodgkin-Huxley equations. That reduction of the Hodgkin-Huxley
equations later became known as the FitzHugh-Nagumo model and has
given us great insight into the mathematical and physiological
complexities of the excitability process. Another consequence of
the Hodgkin-Huxley model, taken to its interpretational extreme,
was the implication that there were microscopic ``channels" in the
membrane through which ions would flow and which were controlled
by membrane potential. There were strong experimental data also
leading to the same conclusion including the binding of
tetrodotoxin (TTX) to nerve membranes to block sodium currents,
titration studies indicating that there were about 20 TTX binding
sites per square micrometer, and membrane noise measurements.
However, it was left to the German physicist Erwin Neher, in
conjunction with the physiologist Bert Sakmann, to develop the
technology and techniques that eventually allowed them to
demonstrate the existence of these ion channels. They were awarded
the Nobel Prize in 1991 for this work. Modifications of the
Hodgkin-Huxley equations were soon proposed for cardiac tissue as
well as a myriad of other excitable cells.

Extensions of the  work of Hodgkin and Huxley soon followed. For
example, J. W. Woodbury (a physicist turned physiologist) and his
student W. E. Crill found that current injected into one cell in a
sheet of heart muscle changed the membrane voltage in nearby cells
in an anisotropic manner. This showed that there must be low
resistance connections between abutting cells in heart tissue and
paved the way for the discovery and characterization of gap
junctions between the cells (in a variety of tissues such as
epithelia). Woodbury also showed that Eyring reaction rate theory,
learned from his famous foster Doctorvater Henry Eyring, can be
used to explain the linear current-voltage relationship of open
sodium channels by choosing the appropriate electrochemical
potential profile encountered by a sodium ion while traversing a
Na ion channel. This, together with other lines of types of
experimental evidence mentioned above, established the feasibility
of the ion channel concept before single channel conductances were
directly measured by Neher.

One of the most remarkable  individuals interested in the dynamic
behavior of simple nervous systems was H. K. Hartline of the John
Hopkins University. Hartline was trained as a physiologist, and
following receipt of his M.D. spent an additional two years at
Hopkins taking mathematics and physics courses. For some
unaccountable reason he was still not satisfied with his training
and obtained funding to study for a further year in Leipzig with
the physicist Werner Heisenberg and a second year in Munich with
Arthur Sommerfeld. Armed with this rather formidable training in
the biological, mathematical, and physical sciences he then
devoted the majority of his professional life at Hopkins to the
experimental study of the physiology of the retina of the
horseshoe crab Limulus. His papers are a marvel of beautiful
experimental work combined with mathematical modelling designed to
explain and codify his findings, and his life work justly earned
him the Nobel Prize in 1967 (with George Wald) ``for his
discoveries concerning the primary physiological and chemical
visual processes in the eye."  As an aside we should point out
that FitzHugh (of the FitzHugh-Nagumo reduction of the
Hodgkin-Huxley model) received his Ph.D. in biophysics (where he
learned mathematics, physics, and chemistry) under Hartline after
completing his biological studies at the University of Colorado.

One can hardly underestimate  the impact that this work in
excitable cell physiology has had on the biological sciences since
the impact is so broad and pervasive. The Notices of the American
Mathematical Society (December 1999) has a very nice article by
Nancy Kopell with some of the mathematical side of the story, and
Nature Neuroscience (November 2000) featured some of this from a
biological perspective in an interesting and lively series of
survey articles.

\subsection*{Molecular biology}

Genetics started in 1866, when  Gregor Mendel first deduced the
basic laws of inheritance. However, modern genetics, with its
capacity to manipulate the very essence of living things, only
came into being with the rise of molecular investigations
culminating in the breakthrough discovery of the structure of DNA;
for which Francis Crick, James D. Watson, and Maurice Wilkins
received the Nobel prize in 1962. The contribution of physics and
physicists to this, what Watson calls Act 1 of molecular biology's
great drama, was seminal. Here we review the work of some of the
physicists who helped shape molecular biology into the exciting
science it currently is.

Max Delbr\"{u}ck received his  doctorate in theoretical physics
from the University of G\"ottingen, and then spent three
postdoctoral years in England, Switzerland, and Denmark. His
interest in biology was aroused during his stay in Denmark by
Niels Bohr's speculation that the complementarity principle of
quantum mechanics might have wide applications to other scientific
fields, and especially to the relation between physics and
biology. Back in Berlin, Delbr\"{u}ck initiated an
interdisciplinary collaboration with Nikolai W. Timofeeff and Karl
G. Zimmer on biologically inspired problems. Based on X-ray
induced mutagenesis experiments and applying concepts from quantum
mechanics, they suggested that chromosomes are nothing more than
large molecules and that mutations can be viewed as ionization
processes. These results were published in 1935. Schr\"{o}dinger's
little book ``What is Life?" (1944) was in part inspired by this
paper.

In 1937, Delbr\"{u}ck moved  from Germany to the United States,
and decided to remain after the start of World War II. At that
time he initiated a fruitful collaboration with Salvador Luria on
the genetic structure of bacteriophage (bacteria-infecting
viruses) and on the genetic mechanism of DNA replication. After
the outbreak of the war, Delbr\"uck and Luria were classified as
'enemy aliens' by the American government despite their open
opposition to the Nazi and Fascist regimes. This classification
fortuitously allowed them to pursue their own investigations
without having to join any military project. For ``their
discoveries concerning the replication mechanism and the genetic
structure of viruses,"  Delbr\"{u}ck and Luria were awarded the
Nobel Prize in 1969, along with Alfred D. Hershey. In the early
1950's Delbr\"{u}ck's research interests shifted from molecular
genetics to sensory physiology, with the goal of clarifying the
molecular nature of the primary transduction processes of sense
organs. Delbr\"{u}ck was also involved in setting up an institute
of molecular genetics at the University of Cologne. It was
formally dedicated on June 22nd, 1962, with Niels Bohr as the
principal speaker. His lecture entitled ``Light and Life
Revisited"  commented on his original one of 1933, which had been
the starting point of Delbr\"{u}ck's interest in biology. It was
to be Bohr's last formal lecture. He died before completing the
manuscript of this lecture for publication.

Erwin  Schr\"{o}dinger is regarded as one of the fathers of
quantum mechanics. However, his interests went far beyond physics.
He was particularly interested in philosophy and biology. Early in
his career, he made substantial contributions to the theory of
color vision. Schr\"{o}dinger's personal life was tumultuous. He
participated as an officer in World War I on the Italian front.
For a variety of reasons, Schr\"{o}dinger moved constantly,
holding positions in Austria, Switzerland, Germany, England, and
then Austria again. Soon after he took up this last position in
Graz, Austria fell into the hands of the Nazis, and
Schr\"{o}dinger escaped to Ireland since his initial departure
from Berlin when the National Socialists took power was considered
an unfriendly act.

In Ireland,  Schr\"{o}dinger joined the Institute for Advanced
Studies in Dublin. His contract required him to give a yearly
series of public lectures. In 1943, he elected to discuss whether
the events in space and time which take place within the spatial
boundary of a living organism can be accounted for by physics and
chemistry in light of the most recent developments in quantum
mechanics and its application to genetics. These lectures were
published in book form in 1944 under the title ``What is Life?"
After discussing how thermodynamics plays a role in the processes
of life and reviewing the not-so-recent results on mutagenesis by
Delbr\"{u}ck et al., Schr\"{o}dinger argued in ``What is Life?"
that life could be thought of in terms of storing and transmitting
information. Chromosomes were thus simply bearers of information.
Because so much information had to be packed into every cell,
Schr\"{o}dinger argued it must be compressed into what he called a
`hereditary code-script' embedded in the molecular fabric of
chromosomes. To understand life, then, it was necessary to
identify these molecules, and crack their code. Schr\"{o}dinger's
book had the very positive effect of popularizing the Delbr\"{u}ck
paper and of rephrasing some important questions derived from it
in a language accessible to the non-expert. The book's publication
could not have been better timed, and it was tremendously
influential. Many of those who would play major roles in the
development of molecular biology were drawn to this field after
reading ``What is is Life?" Schr\"{o}dinger's recruits included
Francis Crick, James D. Watson, Maurice Wilkins, Seymour Benzer,
and Fran\c{c}ois Jacob.

Francis Crick studied  physics at University College, London.
After graduating, he started research for a doctorate, but this
was interrupted by the outbreak of World War II. During the war he
worked as a scientist for the British Admiralty, mainly on
magnetic and acoustic mines. When the war ended, Crick had planned
to stay in military research but, on reading Schr\"{o}dinger's
book, he joined the Medical Research Council Unit in Cambridge to
study biology. In 1951, Crick started a collaboration with James
D. Watson, who came to Cambridge as a postdoctoral fellow. Watson
had originally considered being a naturalist, but he was also
hooked on gene research by Schr\"{o}dinger's book. Linus Pauling
had discovered the alpha helix protein structure by making scale
models of the different parts of the molecule, and working out
possible 3-dimensional schemes to infer which type of helical fold
would be compatible with the underlaying chemical features of the
polypeptide (amino acid) chain. Following Pauling's approach,
Watson and Crick started to look for the structure of DNA, which
in 1944 had been discovered to be the substance making up the
chromosomes. They finally succeeded in the Spring of 1953. Not
only did they determine the structure of DNA, but they also
proposed a scheme for its replication.

Essential for the work  of Watson and Crick were the experimental
results of Rosalind Franklin and Maurice Wilkins. Franklin had a
background in chemistry while Wilkins was a physicist. During
World War II, Wilkins worked in the Manhattan Project. For him, as
for many other of the scientists involved, the actual deployment
of the bombs in Hiroshima and Nagasaki, the culmination of all
their work, was profoundly disillusioning. He considered forsaking
science altogether to become a painter in Paris. However, he too
had read Schr\"{o}dinger's book and biology intervened. Franklin,
working in Wilkins' lab, recorded the DNA X-Ray diffraction
patterns that allowed Watson and Crick to beat Pauling in the race
to determine the structure of DNA. Crick, Watson, and Wilkins
received the Nobel Prize in 1962 ``for their discoveries
concerning the molecular structure of nuclear acids and its
significance for information transfer in living material."
Rosalind Franklin had died at an early age a few years before, and
was not recognized for her essential contributions.

Knowing the structure of DNA was only the start. Next it  was
necessary to find the sequence of genes and chromosomes, to
understand the molecular machinery used to read the messages in
DNA, and to understand the regulatory mechanisms through which the
genes are controlled. These questions were answered by a second
generation of molecular biologists like Seymour Benzer, Sydney
Brenner, Fran\c{c}ois Jacob, Jacques Monod, and Walter Gilbert.
Seymour Benzer and Walter Gilbert had also been educated as
physicists, but were attracted to the excitement of the new
science. Seymour Benzer also heeded the clarion call of the
Schr\"{o}dinger book. He was a pioneer of gene sequencing. Among
other things, Benzer was the first to produce a map of a single
bacteriophage gene, rII, showing how a series of mutations (all
errors in the gene script) were laid out linearly along the viral
DNA.

Walter Gilbert received his doctorate in theoretical  physics and,
after becoming professor at Harvard, worked on particle physics
and quantum field theory for a number of years. Then his interests
shifted. In 1960 Gilbert joined James Watson and Fran\c{c}ois Gros
in a project to identify messenger RNA. After a year of work on
this problem, Gilbert returned to physics only to re-return to
molecular biology shortly afterwards. Some of the more important
contributions of Gilbert and his collaborators to this field are:
the discovery that a single messenger molecule can service many
ribosomes at once and that the growing proteinic chain always
remains attached to a transfer RNA molecule; the isolation of the
lactose repressor, the first example of a genetic control element;
the invention of the rolling circle model, which describes one of
the two ways DNA molecules duplicate themselves; the isolation of
the DNA fragment to which the lac repressor binds; and the
development of rapid chemical DNA sequencing and of recombinant
DNA techniques. Walter Gilbert and Frederick Sanger received the
Nobel Prize in 1980, ``for their contributions concerning the
determination of base sequences in nucleic acids."

\section*{Present and future perspectives}

What we have described so far have been a  few of the significant
advances made in the study of systems in which there was a certain
clear and obvious physics and mathematics component to the
research being carried out. The advances made in the biological
understanding were often quite dependent on the application of
physical and mathematical principles, or the development of the
physics and the mathematics was clearly driven by observations in
biology. This strong interdependence is mirrored in the
highlighting of biologically oriented problems in the new
millennium (January, 2000) issues of Physics Today and the Notices
of the American Mathematical Society as well as the special
November, 2000 Nature Neuroscience issue ``Computational
Approaches to Brain Function." The Notices of the American
Mathematical Society have on several occasions focussed on
problems involving biomathematics (September, 1995) or molecular
biology (April and May, 2002).

Many major universities in  the world have at least one research
group working in these fields. However, listing them all is beyond
the scope or the intent of this article. Our purpose has only been
to illustrate how widespread and important biophysics and
biomathematics have been in the past few centuries and the
increase in their importance in the past few decades.

Darwin's theory states that,  given the environmental conditions,
the fittest individuals are the ones that survive and reproduce.
However, it is impossible to identify the current fittest
individuals whose genes are going to pass to the next generation.
They can be pinpointed only after they have survived. Thus,
according to some, Darwinism is tautological since it only
predicts the survival of the survivors. In trying to foresee the
future of science, we face the same problem. It is not possible to
identify the current areas of scientific research that will play a
relevant role in the development of science and technology. Even
though we acknowledge this problem, it is our belief that, given
the fruitful historical relation and the present blooming of
biological, physical, and mathematical interdisciplinary sciences,
they are going to be so important in the near future that the
avant garde biological scientists will be those with a strong
background in both the biological and the physical-mathematical
sciences.

The mathematical and computational  modelling of biological
systems is a subject of increasingly intense interest. The
accelerating growth of biological knowledge, in concert with a
growing appreciation of the spatial and temporal complexity of
events within cells, tissues, organs, and populations, threatens
to overwhelm our capacity to integrate, understand, and reason
about biology and biological function. The construction, analysis,
and simulation of formal mathematical models is a useful way to
manage such problems. Metabolism, signal transduction, genetic
regulation, circadian rhythms, and various aspects of neurobiology
are just a subset of the phenomena that have been successfully
treated by mathematical modelling. What are the likely areas of
advancement for the future? Predicting the future has fascinated
and confounded man for centuries, probably for as long as he has
been able to articulate the concept of the future. For example,
some relatively recent predictions were:

\begin{quotation}
\noindent
Physics is finished, young man. It's a dead-end street.\\
-Unknown teacher of Max Planck, late 19th century \\[.3cm]
I believe that the motion picture is destined to revolutionize our educational system and that in a few years it will supplant largely, if not entirely, the use of textbooks.\\
-Thomas Edison, 1922\\[.3cm]
It is probable that television drama of high caliber and produced by first-rate artists will materially raise the level of dramatic taste of the nation.\\
-David Sarnoff, 1939
\end{quotation}

Being aware of the almost certain  folly of trying to predict the
future, as illustrated by these quotations, we nevertheless take
the leap and mention several areas in which we feel that
significant advances are likely to take place over the present
century.

\begin{itemize}
\item The sequencing of human and  other genomes has provided a
spectacular amount of data which needs to be organized and
analyzed before its significance becomes clear. The mathematical
techniques necessary to do so are still to be developed. This has
opened a whole new area of research known as bio-informatics,
which is rapidly growing and, presumably, will keep on growing at
an accelerated pace in the next few years. However, we are of the
opinion that the sequence analysis component of bio-informatics
will quickly evolve to become a mere tool widely and easily used
by scientific practitioners (in analogy with the transition from
scientific computing being done on large mainframe computers a few
decades ago, and now being almost exclusively carried out on
inexpensive workstations).

\item The classification aspects  of bio-informatics will be
rapidly replaced by efforts to understand the regulation of gene
networks using established and new techniques from non-linear
dynamics. Mathematical modelling and analysis of the mechanisms of
gene regulation will continue at an ever accelerating pace. This,
in conjunction with the already established ability to produce
``designer" molecular circuits, will be instrumental in the
targeted treatment of disease through gene therapy.

\item  Attempts to understand the noisy interactions in gene
regulation and expression at the single cell level will lead to
the development of new mathematical techniques for dealing with
chemical reactions in which the law of large numbers cannot be
invoked.

\item  The Herculean efforts of countless neurobiologists over the
past century have given us much insight into the functioning of
single neurons as well as the behavior of simple neural circuits
and some extremely simple sensory and motor systems. This progress
will continue and lead to the efficient treatment of many
neuron-related diseases, to a better design of protheses, and
perhaps, to a deeper understanding of the relation between brain
and mind. Shall we, at some time, be able to really understand
phenomena like cognition and memory? Maybe, maybe not. Perhaps, as
some philosophers maintain, the human mind is unable to understand
itself. However, we firmly believe that the neurophysiological
sciences will thrive in the near future, with physics and
mathematics playing a central role in such progress. Examples are
the use of vagal stimulation to abort epileptic seizures, and deep
brain stimulation to control the tremor of Parkinson disease.

\item  Biophysical advances in determining the structure and
dynamic properties of membrane channels and receptors have
proceeded at a rapid pace over the past decade. There is every
reason to anticipate that this will only accelerate in the future.
The accumulated knowledge, in conjunction with modelling and
production of designer molecules, will enable the efficient
development and production of drugs specifically targeted to the
elimination of disease symptoms if not the disease itself.

\item  The accelerated rhythm at which technology is progressing
makes us believe that, in the near future, it will be possible to
combine knowledge and techniques from biology, chemistry,
biochemistry, computer science, engineering and physics to
engineer designer molecules for specific medical and industrial
purposes.

\item  Interdisciplinary work focused in the development of
bio-materials, bio-electronic devices, and bio-mechanical systems
will improve the design of artificial organs, protheses, and
implants through the development of hybrid animate-inanimate
devices.

\item  Epidemiological research aided by mathematical modelling
and statistical analysis will help us understand the dynamics of
disease transmission and to design more efficacious treatment and
vaccination strategies.

\item  The difficulty in collecting high resolution temporal and
spatial data from ecological and meteorological systems has
limited the success of mathematical modelling approaches in these
fields. The availability of more sophisticated geographic
information systems and massive parallel computational power will
alleviate these problems.
\end{itemize}

\section*{Summary}

There  has been a long and rich tradition of fruitful
interdisciplinary interplay between the physical and biological
sciences extending over several centuries, as we have illustrated
with a few examples. Many other examples could have been offered
to illustrate the point, and would simply serve to highlight the
rich interactions between apparently disparate branches of
science. We expect that these interactions and interdependence
will continue and become even stronger in the future.

\section*{Acknowledgments}

We are  grateful to N. Anderson-Mackey, R. FitzHugh and J. Walter
Woodbury for extensive comments on this article, which is largely
based on a lecture with the same title given by MCM 4 May, 2001,
at the University of Oxford as the Leverhulme Professor of
Mathematical Biology for the 2001 academic year. This work was
supported by COFAA-IPN (M{\'e}xico), EDI-IPN (M{\'e}xico), MITACS
(Canada), the Natural Sciences and Engineering Research Council
(NSERC grant OGP-0036920, Canada), and Le Fonds pour la Formation
de Chercheurs et l'Aide ˆ la Recherche (FCAR grant 98ER1057,
Qu{\'e}bec).

\section*{Bibliography}

\begin{itemize}
\item Alessandro Volta web page at University of Pavia. URL: http: //ppp.unipv.it/Volta

\item  Asimov, I. Asimov's Biographical Encyclopedia of Science
and Technology: The Lives and Achievements of 1510 Great
Scientists from Ancient Times to the Present Chronologically
Arranged. Garden City, N.Y.: Doubleday, 1982

\item  Bernal, J. D. Science in History. Cambridge, Mass.: M.I.T.
Press, 1971

\item  Koenigsberger, L. Herman von Helmholtz. New York, N.Y.:
Dover Publications, 1965.

\item  Moore, W. Schr\"{o}dinger, Life and Thought. Cambridge; New
York: Cambridge University Press, 1989

\item  MacTutor History of Mathematics. URL: http:
//www-gap.dcs.st-and.ac.uk/\symbol{126}history

\item Malmivuo, J. and Plonsey, R. Bioelectromagnetism.  New York,
Oxford: University Press, 1995. URL: http:
//butler.cc.tut.fi/\symbol{126}malmivuo/bem/bembook/index.htm

\item Nobel e-Museum. URL: http: //www.nobel.se

\item Scr\"{o}dinger, E. What is Life?:  The Physical Aspects of
the Living Cell. Cambridge: University Press, 1944

\item Watson, J. DNA: The Secret of Life.  New York: Alfred A.
Knopf, 2003
\end{itemize}

\end{document}